\documentclass[aip,graphicx, twocolumn, 10pt, dvipdfmx]{revtex4-1}
\usepackage{graphicx}
\usepackage[dvipdfmx]{color}
\usepackage{bm}
\usepackage[T1]{fontenc}
\draft 

\begin{document}


\title{Spin polarization gate device based on the chirality-induced spin selectivity and robust nonlocal spin polarization} 



\author{Hiroaki Shishido}
 \email[Authors to whom correspondence should be addressed: H. Shishido, ]{shishido@omu.ac.jp.}
\affiliation{Department of Physics and Electronics, Osaka Metropolitan University, Sakai, Osaka 599-8531, Japan
}%
\author{Yuta Hosaka}%
\affiliation{Department of Physics and Electronics, Osaka Prefecture University, Sakai, Osaka 599-8531, Japan
}%

\author{Kenta Monden}
\affiliation{Department of Physics and Electronics, Osaka Prefecture University, Sakai, Osaka 599-8531, Japan
}

\author{Akito Inui}
\affiliation{Department of Physics and Electronics, Osaka Prefecture University, Sakai, Osaka 599-8531, Japan
}

\author{Taisei Sayo}
\affiliation{Department of Physics and Electronics, Osaka Prefecture University, Sakai, Osaka 599-8531, Japan
}%

\author{Yusuke Kousaka}
\affiliation{Department of Physics and Electronics, Osaka Metropolitan University, Sakai, Osaka 599-8531, Japan
}%
\author{Yoshihiko Togawa}
\email[Authors to whom correspondence should be addressed: Y. Togawa, ]{ytogawa@omu.ac.jp.}
\affiliation{Department of Physics and Electronics, Osaka Metropolitan University, Sakai, Osaka 599-8531, Japan
}%
\affiliation{Quantum Research Center for Chirality, Institute for Molecular Science, Okazaki 444-8585, Japan}


\begin{abstract}
Nonlocal spin polarization phenomena are thoroughly investigated in the devices made of chiral metallic single crystals of CrNb$_3$S$_6$ and NbSi$_2$ as well as of polycrystalline NbSi$_2$. We demonstrate that simultaneous injection of charge currents in the opposite ends of the device with the nonlocal setup induces the switching behavior of spin polarization in a controllable manner.  Such a nonlocal spin polarization appears regardless of the difference in the materials and device dimensions, implying that the current injection in the nonlocal configuration splits spin-dependent chemical potentials throughout the chiral crystal even though the current is injected into only a part of the crystal. We show that the proposed model of the spin dependent chemical potentials explains the experimental data successfully.  
The nonlocal double-injection device may offer significant potential to control the spin polarization to large areas because of the nature of long-range nonlocal spin polarization in chiral materials.
 
\end{abstract}


\maketitle 

\section{Introduction}

Conventional semiconductor devices based on electronics are well-established and widely used in modern society. 
For further development, in the last two decades, spin-based technologies, known as spin electronics, have attracted extensive attention as devices with new principles \cite{Pri98, Wol01}.  
In spin electronics devices, charge-to-spin current conversion is a crucial technology. 
Charge-to-spin current conversion is achieved by utilizing the spin Hall effect~\cite{Dya71, Hir99, Zha00}, which was demonstrated initially in paramagnetic metals \cite{Val06} and now available even in topologically protected surface spin currents in topological insulators~\cite{Li14, Shi14, Mel14, Fan14, And14, Deo14}.

Chirality-induced spin selectivity (CISS) is another potential method for the charge-to-spin current conversion, where electrons flowing through a chiral material are spin-polarized reflecting the handedness of the material. 
The CISS phenomena were initially reported in chiral molecules via spin-polarized photocurrent emission \cite{Goh11} and tunneling transport experiments \cite{Xie11}. 
Recent studies have shown that the spin polarization can be induced by the charge current due to the CISS effect in inorganic metals, such as in a paramagnetic state of chiral helimagnet CrNb$_3$S$_6$~\cite{Inu20, Nab20}. 
It was observed that the spin polarization is dependent on the handedness of the crystals and is protected without charge current over 1\,$\mu$m in a nonlocal measurement configuration~\cite{Inu20}.
This record of the spin polarization length was updated to 10\,$\mu$m by measuring the CISS effect in nonmagnetic chiral crystals NbSi$_2$ and TaSi$_2$ \cite{Shi21}.
Moreover, the CISS effect has been confirmed in polycrystalline samples of NbSi$_2$ and TaSi$_2$. 
The spin polarization was observed even a few millimeters away from the charge current injection site~\cite{Shis21}.  
The record of the spin polarization length is summarized in the literature~\cite{YT23}.
Furthermore, it has been reported that the CISS effect is observed in single crystalline NbSi$_2$ samples of more than 6\,cm in length~\cite{Kou23}.

In this paper, we demonstrate the spin polarization response induced by simultaneous nonlocal injection of charge currents from both sides of the crystal in micrometer-sized single and millimeter-sized polycrystalline chiral crystals.
The superposition of spin polarization can be explained by the model that the current injection with the nonlocal setup induces the splitting of spin-dependent chemical potentials throughout the chiral crystal even when the current is injected into only a part of the sample. 
The controllability of spin polarization to large areas is successfully achieved in the nonlocal double-injection device made of chiral material.

\section{Experimental methods}

\begin{figure}[th]
\includegraphics[width=0.9\linewidth]{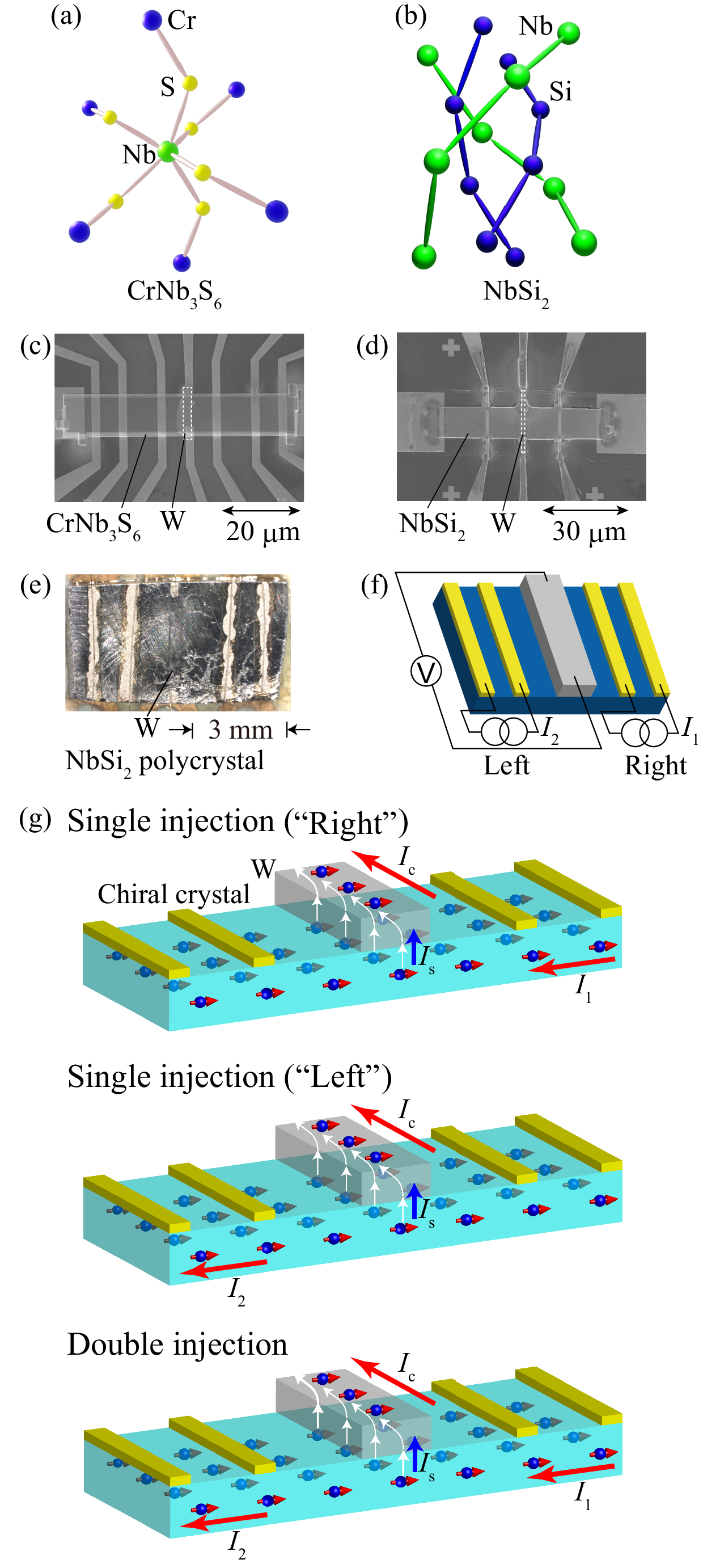}
\caption{Crystal structures of CrNb$_3$S$_6$ (a) and NbSi$_2$ (b). 
SEM images of the devices fabricated from single crystals of CrNb$_3$S$_6$ (c) and NbSi$_2$ (d). (e) The optical photograph of a device fabricated from NbSi$_2$ polycrystalline samples. (f) A schematic of the device structure and measurement configuration. The mechanism of device operation is schematically drawn in (g).}
\label{Crystal_SEM}
\end{figure}

The hexagonal monoaxial chiral crystals of CrNb$_3$S$_6$ and NbSi$_2$ belong to the space group $P6_322$~(No. 182) and $P6_222$~(No. 180) or $P6_422$~(No. 181), respectively.
In CrNb$_3$S$_6$, Cr ions are intercalated between NbSi$_2$ layers resulting in a left- or right-handed crystal structure with lattice constants of $a$ = 0.57\,nm and $c$ = 1.21\,nm, as illustrated in Fig.~\ref{Crystal_SEM}(a).
NbSi$_2$ exhibits a C40-type crystal structure with lattice constants of $a$ = 0.4798\,nm and $c$ = 0.6592\,nm, as shown in Fig.~\ref{Crystal_SEM}(b).
CrNb$_3$S$_6$ shows monoaxial chiral helimagnetism below 132\,K \cite{Miy83, Tog12, Tog_Rev16}, whereas NbSi$_2$ exhibits diamagnetism and no magnetic order \cite{Got93}.

Single crystals of CrNb$_3$S$_6$ and NbSi$_2$ were grown using chemical vapor transport \cite{Kou09} and laser floating zone method \cite{Kou23}, respectively.
For the CISS measurements, samples were fabricated by using focused ion beam milling with dimensions of 60.8\,$\mu$m (49.1\,$\mu$m) in length, 11.2\,$\mu$m (10.5\,$\mu$m) in width, and 1.0\,$\mu$m (1.0\,$\mu$m) in thickness for CrNb$_3$S$_6$ (NbSi$_2$), as shown in Fig.~\ref{Crystal_SEM}(c) [Fig.~\ref{Crystal_SEM}(d)]. 
Polycrystalline ingots of NbSi$_2$ were arc-melted in an Ar atmosphere from compound powders.
Samples were cut from the bulk ingots with typical dimensions of several millimeters in length and width and of 1\,mm in depth, as shown in Fig.~\ref{Crystal_SEM}(e).

In the present study, all the CISS measurements were performed at room temperature without applying any magnetic fields, ensuring that the contribution of the conventional Hall effect was excluded.  
Figure~\ref{Crystal_SEM}(f) shows the schematic of the device structure and CISS measurements.
Four gold electrodes and a tungsten (W) electrode with a thickness of 6\,nm are deposited on the chiral crystals. 
Charge currents $I_1$ and $I_2$ are applied into the ``right'' and ``left'' 
regions, respectively, which induce the spin polarization due to the CISS response, as schematically drawn in Fig.~\ref{Crystal_SEM}(g). 
The spin current $I_{\rm s}$ flows into the W electrode due to the difference in spin-dependent chemical potential between the chiral sample and the W electrode. 
The spin current in the W electrode is then converted to a transverse electrical current $I_{\rm c}$ via the inverse spin Hall effect (SHE)~\cite{Val06, Sai06, Kim07}. 
Note that NbSi$_2$ is a diamagnetic metal and thus has no localized spins. CrNb$_3$S$_6$ is in a paramagnetic state at room temperature. Therefore, the present transverse voltage is not induced by the anomalous Hall effect (AHE) but by the inverse SHE via the CISS. This is in contrast with the previous study using chiral metallo-bio-organic crystals, where a transverse voltage is induced by the AHE due to the ferromagnetic property of the crystals~\cite{Gor21}.

\section{Experimental results and discussion}

\subsection{Nonlocal CISS signals with conventional setup (for single injection of charge current)}

\begin{figure}[t]
\includegraphics[width=0.9\linewidth]{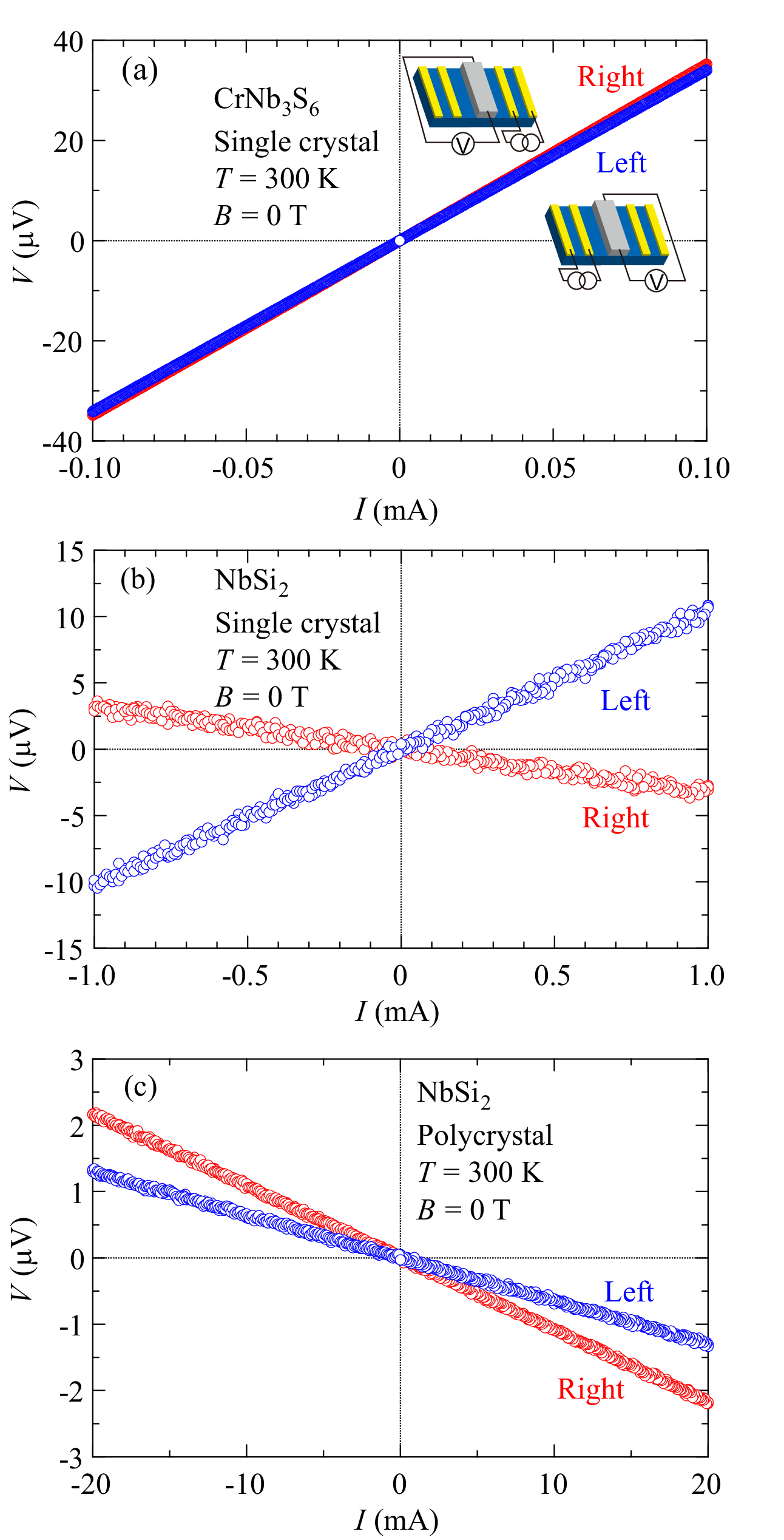}
\caption{The nonlocal CISS signals for (a) CrNb$_3$S$_6$ single crystals, (b) NbSi$_2$ single crystals, and (c) bulk polycrystalline NbSi$_2$.
All the measurements were performed at room temperature without magnetic fields \cite{}.
}
\label{V_xy}
\end{figure}

Figure~\ref{V_xy} displays the CISS signal $V$ as a function of the applied current $I$ using the nonlocal setup.  
The linear $I$--$V$ characteristics of the nonlocal CISS signals were observed not only in the micrometer-sized single crystals of CrNb$_3$S$_6$ [Fig.~\ref{V_xy}(a)] and NbSi$_2$ [Fig.~\ref{V_xy}(b)] but also in the bulk polycrystalline NbSi$_2$ [Fig.~\ref{V_xy}(c)], as previously reported in other publications~\cite{Inu20, Shi21, Shis21}.

Notably, the distance between current and voltage contacts of bulk polycrystalline samples is 1.5\,mm, which is two order of magnitude longer than those of micrometer-sized single crystals. 
The $V/I$ slope is 0.3 $\Omega$ in the single crystal CrNb$_3$S$_6$, while it is about 3 m$\Omega$ and 0.1 m$\Omega$ in the single and polycrystal NbSi$_2$, respectively. 
These slope intensities are of the same order as those reported in previous studies~\cite{Inu20, Shi21, Shis21}, suggesting a successful observation of the CISS phenomena.

The sign of the slope is determined by the handedness at the current injected region of chiral crystals~\cite{Inu20, Shi21}. 
Observing the same signs in the ``left'' and ``right'' sides for the CrNb$_3$S$_6$ and polycrystalline NbSi$_2$ indicates the same handedness in both regions. 
In contrast, the opposite signs observed in the single crystalline NbSi$_2$ denote the opposite handedness in the ``left'' and ``right'' regions.

\subsection{Nonlocal CISS signals for double injection of charge currents with nonlocal setup}
 
\begin{figure}[b]
\includegraphics[width=0.9\linewidth]{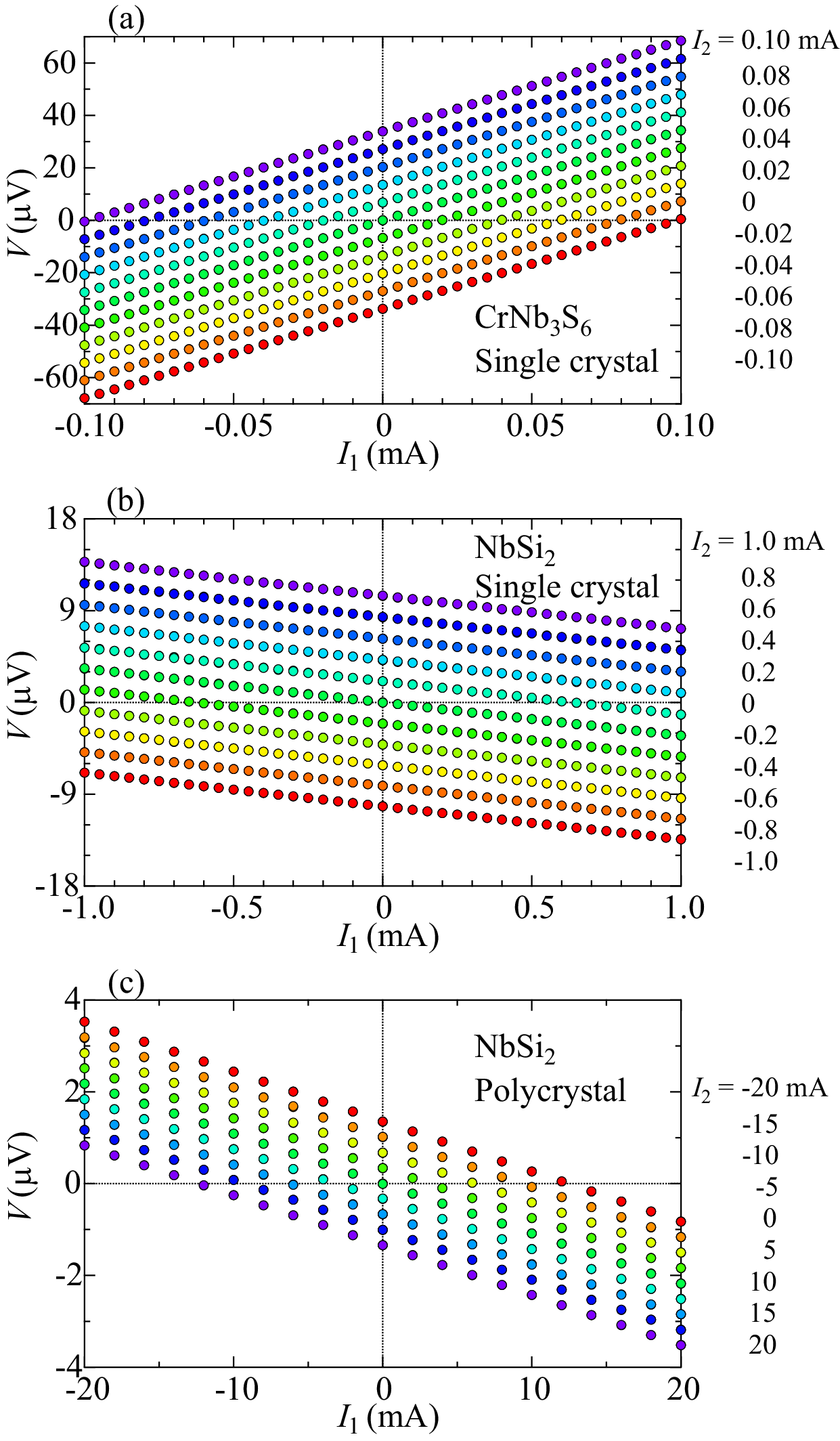}

\caption{
A dataset of $I$--$V$ characteristics is provided for the charge current ($I_1$) injected into the ``right'' region for (a) single crystals of CrNb$_3$S$_6$, (b) single crystals of NbSi$_{2}$, and (c) polycrystalline NbSi$_{2}$, under various injection charge currents $I_2$ into the ``left'' region.
}
\label{double_inp}
\end{figure}

When charge currents are simultaneously injected into the ``left'' and ``right'' regions, the CISS signal appears as a superposition of the CISS signals, separately induced by the current injection into each regions.   
Figure~\ref{double_inp} shows that the linearity of the $I$--$V$ curves with regard to the current ($I_1$) injected into the ``right'' region is maintained even when the additional current ($I_2$) is injected into the ``left'' region. 
In addition, the $I$--$V$ curves shift upward or downward as a function of $I_2$ injected into the ``left'' region. 
The polarity and intensity of the shifts are consistent with those expected from the $I$--$V$ curves for the single injection of charge current, shown in Fig.~\ref{V_xy}.

\begin{figure}[t]
\includegraphics[width=0.9\linewidth]{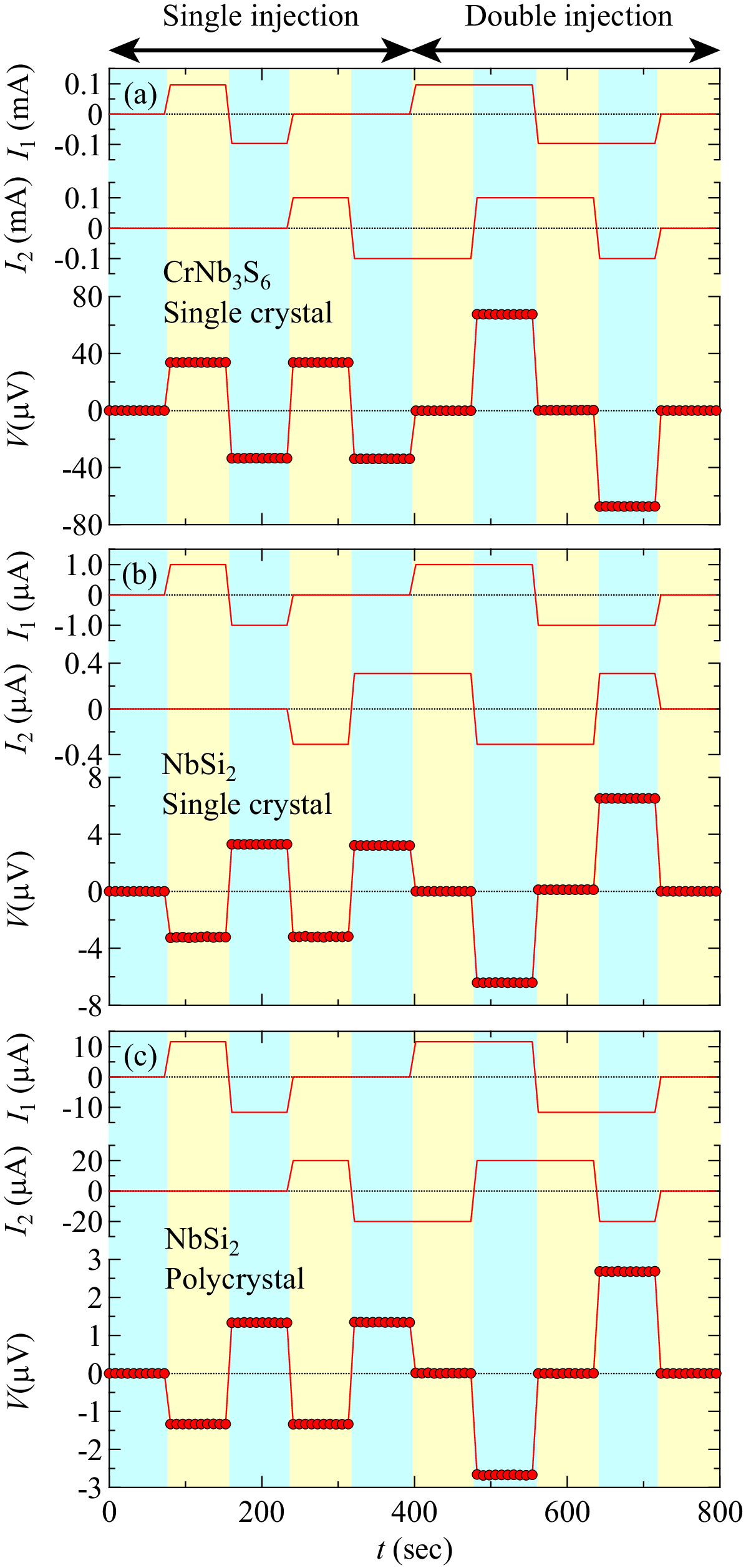}
\caption{Dependence of the CISS signals on the double injection of charge currents with the nonlocal setup for CrNb$_3$S$_6$ single crystals (a), NbSi$_2$ single crystals (b), and polycrystalline NbSi$_2$ (c). The input charge currents were injected into the ``right'' ($I_1$) and ``left'' ($I_2$) regions.  First five columns in the panels represent the single injection of input current, while the latter five columns correspond to the operation of double injection. 
}
\label{logic}
\end{figure}

\begin{figure}[t]
\includegraphics[width=0.9\linewidth]{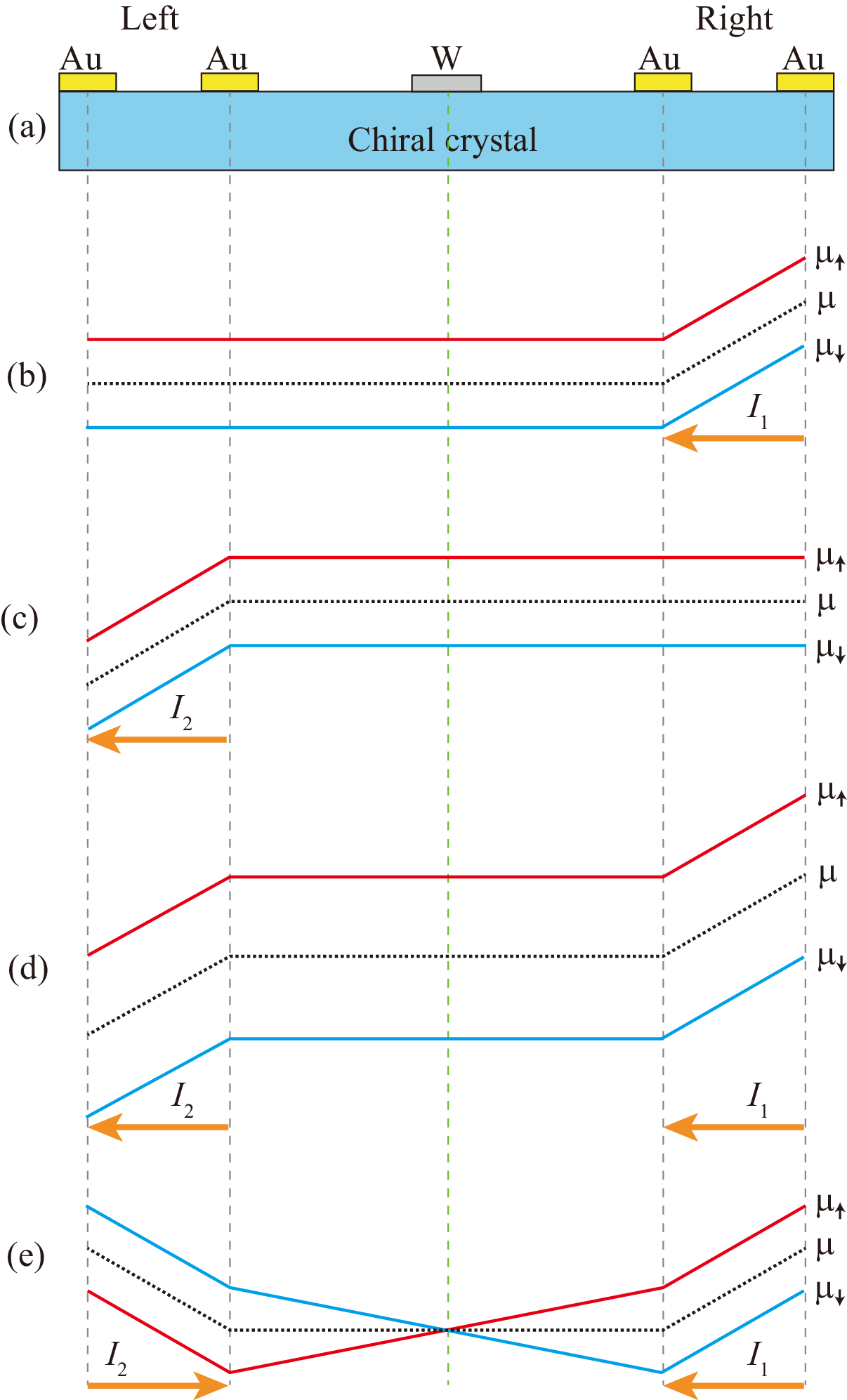}
\caption{Schematics of the spin-dependent chemical potentials induced by the double injection of charge currents in the nonlocal configuration. (a) Side view of a CISS device for the double injection. The distribution of spin-dependent chemical potentials triggered by the single injection of charge current on the ``right'' (b) and ``left'' regions (c). The double injection of charge currents flowing in the same direction gives the same polarity of the CISS signals and, thus, enhances the potential difference at the center region, as shown in (d). On the other hand, the charge currents with the opposite direction in the nonlocal setup induce the opposite polarity of the CISS signals and, thus, the potential difference intersects at the center, as shown in (e).     
}
\label{chemP}
\end{figure}


On the basis of the $I$--$V$ characteristics for the double injection of charge currents with the nonlocal setup, we demonstrate that the CISS output is controllable in terms of the polarity and intensity of the signal.
Figure~\ref{logic} indeed shows such switching behavior of the CISS signal against the input charge currents $I_1$ and $I_2$. 

For the single injection of input current into the single crystal of CrNb$_3$S$_6$ [see the first five columns in Fig.~\ref{logic}(a)], the CISS signal becomes 33.8 and $-$33.5\,$\mu$V for $I_1$ = 0.96 and $-$0.96\,mA, respectively, with $I_2$ being 0\,mA. Alternatively, the input current of $I_2$ = 1.00 and $-$1.00\,mA with $I_1$ fixed to be 0\,mA results in the CISS signal of 33.7 and $-$33.8\,$\mu$V, respectively, the intensities of which are almost as large as the those obtained for the former case (by the $I_1$ injection).

Next, let us see the operation of double injection, as shown in the latter five columns in Fig.~\ref{logic}(a). The CISS signal is suppressed to be less than 0.3\,$\mu$V for the simultaneous injection of input currents ($I_1, I_2$) = (0.96, -1.00) or (-0.96, 1.00), given in the unit of mA. On the other hand, the CISS signal reaches 67.5\, and $-$67.3\,$\mu$V for the double injection of ($I_1, I_2$) = (0.96, 1.00) and (-0.96, -1.00), respectively. 

Similar switching behavior of the CISS signals is also obtained for the double injection of charge currents with the nonlocal setup in single crystals and bulk polycrystals of NbSi$_2$, as shown in Figs.~\ref{logic}(b) and~\ref{logic}(c).

We deduce the spin-dependent chemical potentials across the CISS device from the double-injection CISS response. First, injecting the charge currents in the nonlocal configuration induces the CISS effect, which splits the spin-dependent chemical potentials. 
Importantly, such a split between the up-spin chemical potential $\mu_\uparrow$ and down-spin chemical potential $\mu_\downarrow$ should be transmitted over the chiral crystal without any additional local current, as shown in Figs.~\ref{chemP}(b) and ~\ref{chemP}(c). This picture is supported by the long-range spin polarization observed in the nonlocal measurements. The direction of the split depends on the direction of the injected current and handedness of the crystal. Here, for simplicity, the device is assumed to be made of the enantio-pure crystal.

When the charge currents with the same polarity are injected into both sides of the crystal, as illustrated in Fig.~\ref{chemP}(d), the splits between $\mu_\uparrow$ and $\mu_\downarrow$ are added over the crystal and thus the intensified CISS signal is detected at the center of the device, where the W electrode is deposited. 
Conversely, when the polarities of the injected charge currents are opposite to each other, $\mu_\uparrow$ and $\mu_\downarrow$ intersect at the center of the device, as shown in Fig.~\ref{chemP}(e). 
Consequently, no CISS signal is detected since $\mu_\uparrow$ and $\mu_\downarrow$ degenerate at the center of the device. 
Note that the split between $\mu_\uparrow$ and $\mu_\downarrow$ may vary along the device, which could potentially induce the potential difference even at the center region and might generate the spin current into the W electrode. 
Interestingly, up-spin electrons flow leftward in the chiral material, while down-spin electrons flow rightward because of a spatial gradient of the chemical potential. In particular, a pure spin current flows even in the chiral material in this situation.

\begin{figure}[t]
\includegraphics[width=0.9\linewidth]{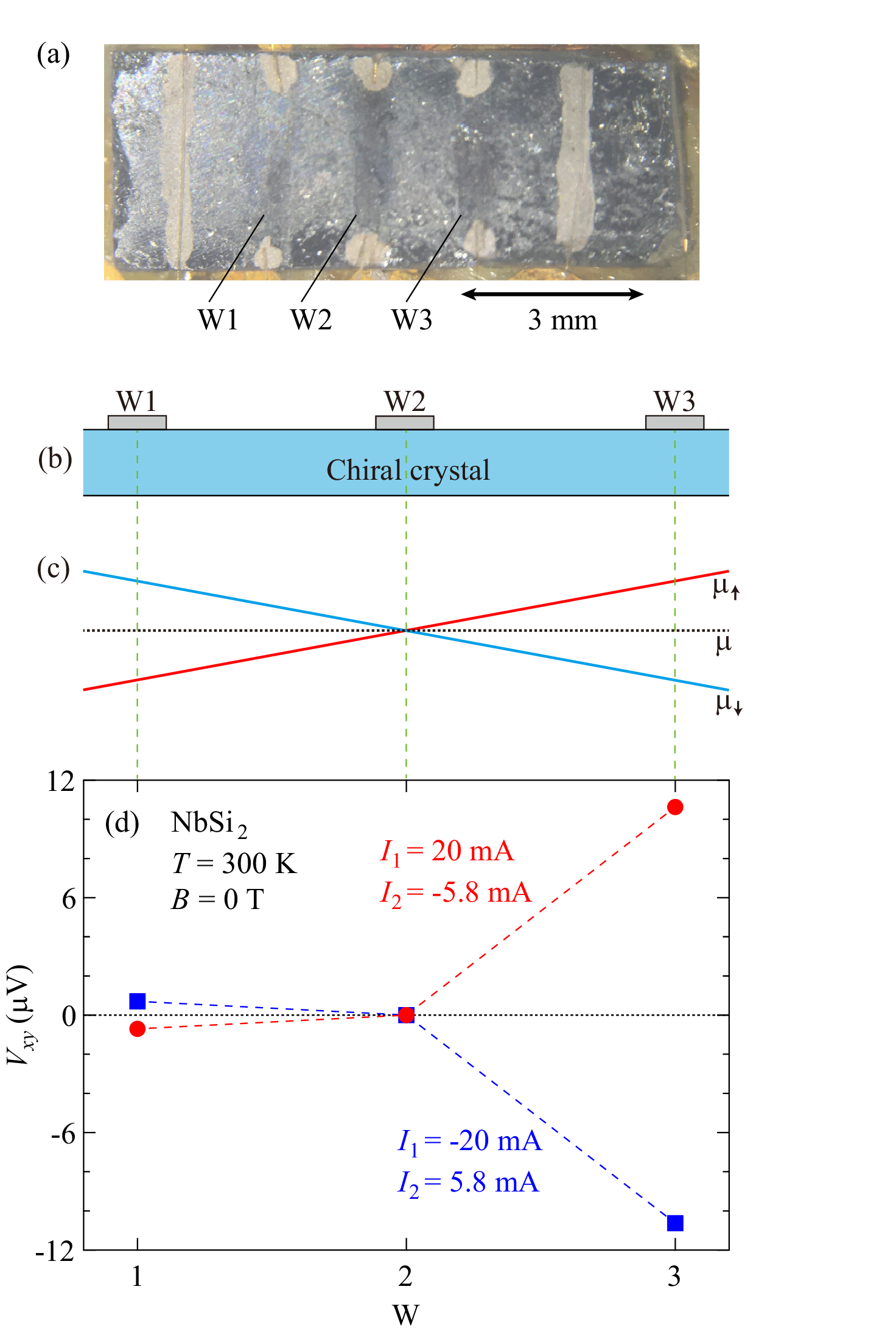}
\caption{(a) The optical photograph of the double-injection CISS device with the three W electrodes. (b) A schematic of the side-view of the device and (c) the distribution of spin-dependent chemical potentials along the chiral crystal. (d) The nonlocal CISS signals detected at each W electrode for the double injection of charge currents.       
}
\label{tripleW}
\end{figure}

To confirm the validity of our spin-dependent chemical potentials model, a spatial distribution of the CISS signals was examined in a bulk NbSi$_2$ double-injection CISS device with the three W electrodes, as shown in Fig.~\ref{tripleW}. 
It is expected that, when the charge currents in the nonlocal configuration with the opposite polarity are applied, $\mu_\uparrow$ and $\mu_\downarrow$ degenerate at the center of the device, resulting in the disappearance of the CISS signal at the center (W2) electrode. 
However, the difference between $\mu_\uparrow$ and $\mu_\downarrow$ should remain finite and opposite in the right and left regions distant from the center. 
Therefore, finite CISS signals with the opposite polarity could be detected at the left (W1) and right (W3) electrodes, as shown in Fig.~\ref{tripleW}(c).

The CISS signals with the input currents ($I_1$, $I_2$) = (20, -5.8) and (-20, 5.8), in the unit of mA, are presented in Fig.~\ref{tripleW}(d). 
A finite CISS signal of $-$0.70 and 0.70\,$\mu$V respectively appears at the W1 electrode, whereas that of 10.6 and $-$10.6\,$\mu$V is observed at the W3 electrode. 
In contrast, the CISS signals at the W2 electrode are less than 2\,nV in both cases. 
Precisely speaking, in the experiments, the combination of $I_1$ and $I_2$ values was tuned to minimize the CISS signal at the W2 electrode. 
Then, the CISS signals at the W1 and W3 electrodes were monitored.
The present observations are qualitatively consistent with our expectation based on the spin-dependent chemical potential model, which predicts finite CISS signals with the opposite polarity at the W1 and W3 electrodes and no CISS signal at the W2 electrode.

\subsection{Absorption of spin polarization by W electrodes (for single injection of charge current)}

\begin{figure}[t]
\includegraphics[width=0.9\linewidth]{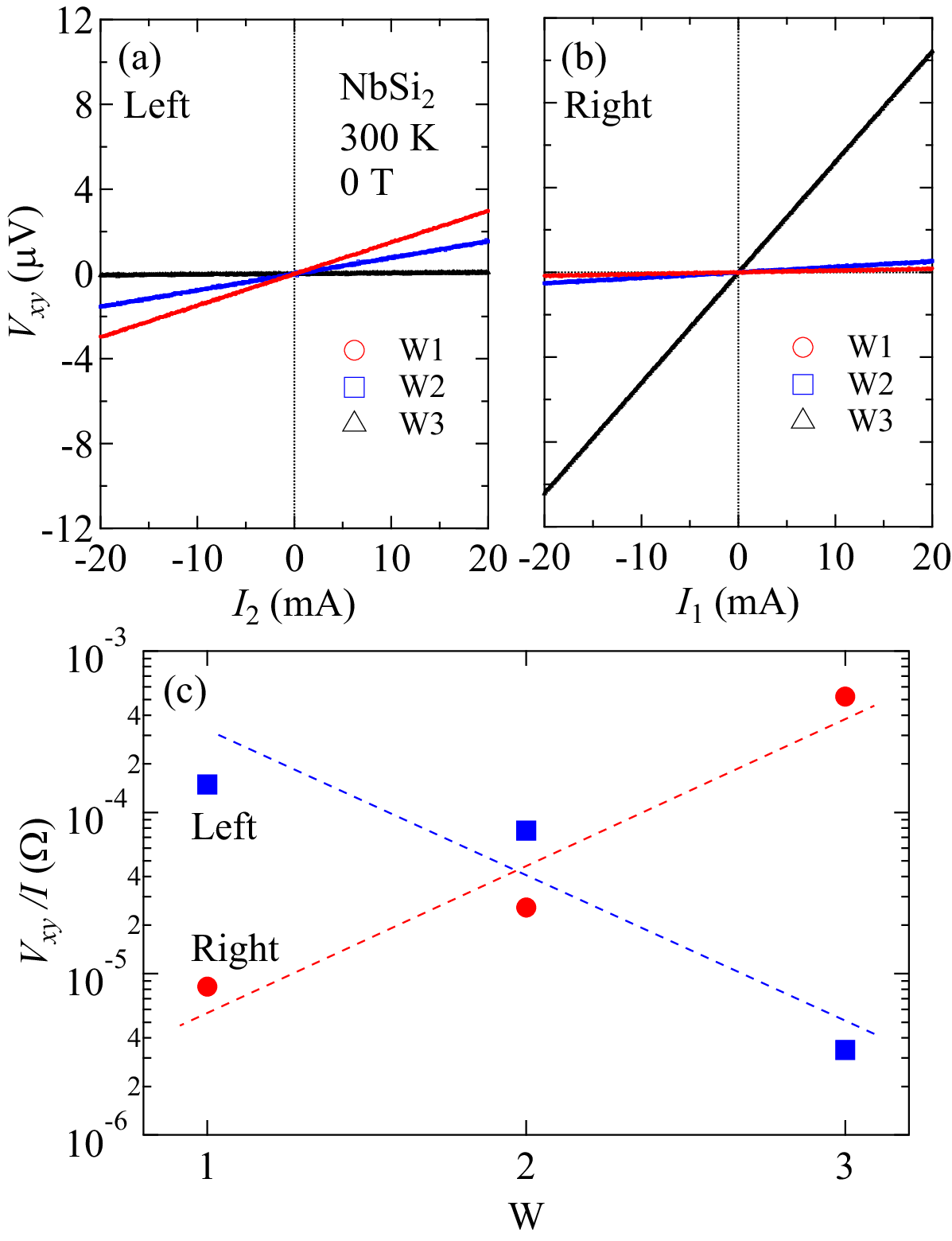}
\caption{The CISS signals at the three W electrodes with the single injection of charge current into  ``left'' (a) and ``right'' (b) regions. The slope $V_{xy}$/$I$ is summarized in (c), where the dashed lines are provided as a guide to the eyes.}
\label{absorption}
\end{figure}

The data in Fig.~\ref{tripleW}(d) shows the difference in the CISS intensity between the W1 and W3 electrodes, which may indicate some features beyond the naive picture given in Fig.~\ref{tripleW}(c). 
In particular, the presence of additional W electrodes may cause a reduction in the CISS signal.

To see the influence of the additional W electrodes, the nonlocal CISS signals were examined 
in the same device with the three W electrodes as that used in Fig.~\ref{tripleW}. 
Figure~\ref{absorption} shows that the signal intensity gradually decreases from the left to the right when the current is injected into the ``left'' region in the single injection of charge current. Conversely, when the current is injected into the ``right'' region, the signal intensity displays a steep increase from the left to the right. 
With an increase in the number of W electrodes in the direction away from the current injection region, the nonlocal CISS signal appears to decrease. 
Furthermore, this result indicates that the split of spin-dependent chemical potentials reduces by the absorption of the spin current into the W electrode. 

\section{Summary}
The nonlocal CISS measurements were thoroughly performed in the double-injection CISS devices made of several chiral materials with different dimensions. The data shows qualitatively similar response of spin polarization, regardless of the difference in materials and device sizes. Furthermore, it is demonstrated that the CISS response could be identified by using the model of spin-dependent chemical potentials, which is derived from the long-range spin polarization in chiral materials. In addition, we argue the possibility that the spin polarization decreases each time when the spin current is absorbed into the W electrode.

The double-injection CISS device may offer significant potential to revolutionize the spin polarization response in the electronic devices. Importantly, these devices can work over a wide range of dimensions due to the nature of long-range spin polarization and operate on the basis of spin polarization rather than electron (charge) flow or accumulation, which ensures a large reduction of heat generation caused by Joule heating. A wide range of chiral materials could be utilized for fabricating the spin polarization devices.

\section*{Supplementary Material}
The supplementary material shows a temperature dependence of the CISS signals in the NbSi$_2$ polycrystalline device used in Figs.~\ref{tripleW} and~\ref{absorption}. 
A different temperature dependence was obtained in the transverse and longitudinal signals, indicating that the present transverse signal is not due to misalignments of the detection electrodes but to the intrinsic CISS effect.

\begin{acknowledgments}
We sincerely thank Yusuke Kato and Hiroaki Kusunose for fruitful discussions. We acknowledge support from Grants-in-Aid for Scientific Research (Grant Nos. 17H02767, 17H02923, 21H01032, 22H01944, and 23H00091) and the Research Grant of Specially Promoted Research Program by Toyota RIKEN, and the Joint Research by Institute for Molecular 
Science (IMS program No. 23IMS1101). 

\end{acknowledgments}

\section*{Data Availability Statement}
The data supporting the findings of the present study is available from the corresponding authors upon request.

\nocite{*}


\begin{thebibliography}{99}\label{sec:TeXbooks}%


\bibitem{Pri98}
G. A. Prinz, Magnetoelectronics, Science {\bf 282}, 1660-1663 (1998).
\bibitem{Wol01}
S. A. Wolf. D. Awschalomr, A. Buhrmanj, M. Daughtons, S. V. Moln\'arm, L. Roukesa, Y. Chtchelkanovaand, and D. M. Treger, Spintronics: A spin-based electronics vision for the future, Science {\bf 294}, 1488-1495 (2001).

\bibitem{Dya71} 
M. I. Dyakonov, and V. I. Perel, Current-induced spin orientation of electrons in semiconductors, Phys. Lett. {\bf 35}, 459-460 (1971).
\bibitem{Hir99} 
J. E. Hirsch, Spin Hall effect, Phys. Rev. Lett. {\bf 83}, 1834-1837 (1999).
\bibitem{Zha00} 
S. Zhang, Spin Hall effect in the presence of spin diffusion, Phys. Rev. Lett. {\bf 85}, 393-396 (2000).


\bibitem{Val06}
S. O. Valenzuela and M. Tinkham, Direct electronic measurement of the spin Hall effect,  Naure \textbf{442}, 176 (2006).

\bibitem{Shi14} 
Y. Shiomi, K. Nomura, Y. Kajiwara, K. Eto, M. Novak, Kouji Segawa, Yoichi Ando, and E. Saitoh, Spin-electricity conversion induced by spin injection into topological insulators, Phys. Rev. Lett. {\bf 113}, 196601 (2014).


\bibitem{Li14}
C. H. Li, O. M. J. van$^{\prime}$tErve, J. T. Robinson, Y. Liu, L. Li, and B. T. Jonker, Electrical detection of charge-current-induced spin polarization due to spin-momentum locking in Bi$_2$Se$_3$,  Nat. Nanotechnol. {\bf 9}, 218-224 (2014). 

\bibitem{Mel14}
A. R. Mellnik, J. S. Lee, A. Richardella, J. L. Grab, P. J. Mintun, M. H. Fischer, A. Vaezi, A. Manchon, E.-A. Kim, N. Samarth, and D. C. Ralph, Spin-transfer torque generated by a topological insulator, Nature {\bf 511}, 449-451 (2014). 
\bibitem{Fan14}
Y. Fan, P. Upadhyaya, X. Kou, M. Lang, S. Takei, Z. Wang, J. Tang, L. He, L.-T. Chang, M. Montazeri, G. Yu, W. Jiang, T. Nie, R. N. Schwartz, Y. Tserkovnyak, and K. L. Wang, Magnetization switching through giant spin-orbit torque in a magnetically doped topological insulator heterostructure, Nat. Mater. {\bf 13}, 699-704 (2014). 
\bibitem{And14}
Y. Ando, T. Hamasaki, T. Kurokawa, K. Ichiba, F. Yang, M. Novak, S. Sasaki, K. Segawa, Y. Ando, and M. Shiraishi, Electrical detection of the spin polarization due to charge flow in the surface state of the topological insulator Bi$_{1.5}$Sb$_{0.5}$Te$_{1.7}$Se$_{1.3}$, Nano Lett. {\bf 14}, 6226-6230 (2014).
\bibitem{Deo14}
P. Deorani, J. Son, K. Banerjee, N. Koirala, M. Brahlek, S. Oh, and H. Yang, Observation of inverse spin Hall effect in bismuth selenide, Phys. Rev. B {\bf 90}, 094403 (2014).



\bibitem{Goh11} 
B. G\"{o}hler, V. Hamelbeck, T. Z. Markus, M. Kettner, G. F. Hanne, Z. Vager, R. Naaman, H. Zacharias, Spin selectivity in electron transmission through self-assembled monolayers of double-stranded DNA, Science \textbf{331}, 894 (2011).
\bibitem{Xie11}
Z. Xie, T. Z. Markus, S. R. Cohen, Z. Vager, R. Gutierrez, R. Naaman, Spin specific electron conduction through DNA oligomers, Nano Lett. \textbf{11}, 4652 (2011).

\bibitem{Inu20}
A. Inui, R. Aoki, Y. Nishiue, K. Shiota, Y. Kousaka, H. Shishido, D. Hirobe, M. Suda, J. Ohe, J. Kishine, H. Yamamoto, Y. Togawa, Chirality-induced spin-polarized state of a chiral crystal CrNb$_3$S$_6$, Phys. Rev. Lett. \textbf{124}, 16602 (2020).
\bibitem{Nab20}
Y. Nabei, D. Hirobe, Y. Shimamoto, K. Shiota, A. Inui, Y. Kousaka, Y. Togawa, and H. M. Yamamoto, Current-induced bulk magnetization of a chiral crystal CrNb$_3$S$_6$, Appl. Phys. Lett. \textbf{117}, 052408 (2020).


\bibitem{Shi21}
K. Shiota, A. Inui, Y. Hosaka, R. Amano, Y. \={O}nuki, M. Hedo, T. Nakama, D. Hirobe, J. Ohe, J. Kishine, H. M. Yamamoto, H. Shishido, and Y. Togawa, Chirality-induced spin polarization over macroscopic distances in chiral disilicide crystals, Phys. Rev. Lett. \textbf{127}, 126602 (2021). 

\bibitem{Shis21}
H. Shishido, R. Sakai, Y. Hosaka, Y. Togawa Appl. Detection of chirality-induced spin polarization over millimeters in polycrystalline bulk samples of chiral disilicides NbSi$_2$ and TaSi$_2$, Phys. Lett. \textbf{119}, 182403 (2021).

\bibitem{YT23}
Y. Togawa, A. S. Ovchinnikov, and J. Kishine, Generalized Dzyaloshinskii-Moriya interaction and chirality-induced phenomena in chiral crystals, J. Phys. Soc. Jpn. \textbf{92}, 081006 (2023).

\bibitem{Kou23}
Y. Kousaka, T. Sayo, S. Iwasaki, R. Saki, C. Shimada, H. Shishido, and Y. Togawa, Chirality-selected crystal growth and spin polarization over centimeters of transition metal disilicide crystals, Jpn. J. Appl. Phys. {\bf 62}, 015506 (2023).

\bibitem{Miy83}
T. Miyadai, K. Kikuchi, H. Kondo, S. Sakka, M. Arai, and Y. Ishikawa, Magnetic properties of Cr$_{1/3}$NbS$_2$, J. Phys. Soc. Jpn. \textbf{52}, 1394 (1983).

\bibitem{Tog12}
Y. Togawa, T. Koyama, K. Takayanagi, S. Mori, Y. Kousaka, J. Akimitsu, S. Nishihara, K. Inoue, A. S. Ovchinnikov, and J. Kishine, Chiral magnetic soliton lattice on a chiral helimagnet, Phys. Rev. Lett. \textbf{108}, 107202 (2012).

\bibitem{Tog_Rev16}
Y. Togawa, Y. Kousaka, K. Inoue, and J. Kishine, 
Symmetry, structure, and dynamics of monoaxial chiral magnets, J. Phys. Soc. Jpn. \textbf{85}, 112001 (2016).

\bibitem{Got93}
U. Gottlieb, A. Sulpice, R. Madar, and O. Laborde, Magnetic susceptibilities of VSi$_2$, NbSi$_2$ and TaSi$_2$ single crystals, J. Phys.: Condens. Matter {\bf 5}, 8755 (1993).

\bibitem{Kou09}
Y. Kousaka, Y. Nakao, J. Kishine, M. Akita, K. Inoue, and J. Akimitsu, Chiral helimagnetism in T$_{1/3}$NbS$_2$ (T=Cr and Mn), Nucl. Instrm. Methods Phys. Res., Sect. A {\bf 600}, 250 (2009).

\bibitem{Sai06}
E. Saitoh, M. Ueda, H. Miyajima, G. Tatara, Conversion of spin current into charge current at room temperature: Inverse spin-Hall effect, Appl. Phys. Lett. \textbf{88}, 182509 (2006).
\bibitem{Kim07}
T. Kimura, Y. Otani, T. Sato, S. Takahashi, S. Maekawa, Room-temperature reversible spin Hall effect, Phys. Rev. Lett. \textbf{98}, 156601 (2007).

\bibitem{Gor21}
N. Goren, T. K. Das, N. Brown, S. Gilead, S. Yochelis, E. Gazit, R. Naaman, Y. Paltiel, Metal organic spin transistor, Nano Lett. \textbf{21}, 8657 (2021).


\end{thebibliography}


\pagebreak
\widetext
\begin{center}
\textbf{\large Supplemental Materials for \\
``Spin polarization gate device based on the chirality-induced spin selectivity and robust nonlocal spin polarization''}
\end{center}

\setcounter{equation}{0}
\setcounter{figure}{0}
\setcounter{table}{0}
\setcounter{page}{1}
\makeatletter

\renewcommand{\figurename}{FIG. S\hspace{-0.8mm}}
\renewcommand{\bibnumfmt}[1]{[S#1]}
\renewcommand{\citenumfont}[1]{S#1}

\section{Temperature dependence of CISS signals}

\begin{figure}[h]
\includegraphics[width=0.5\linewidth]{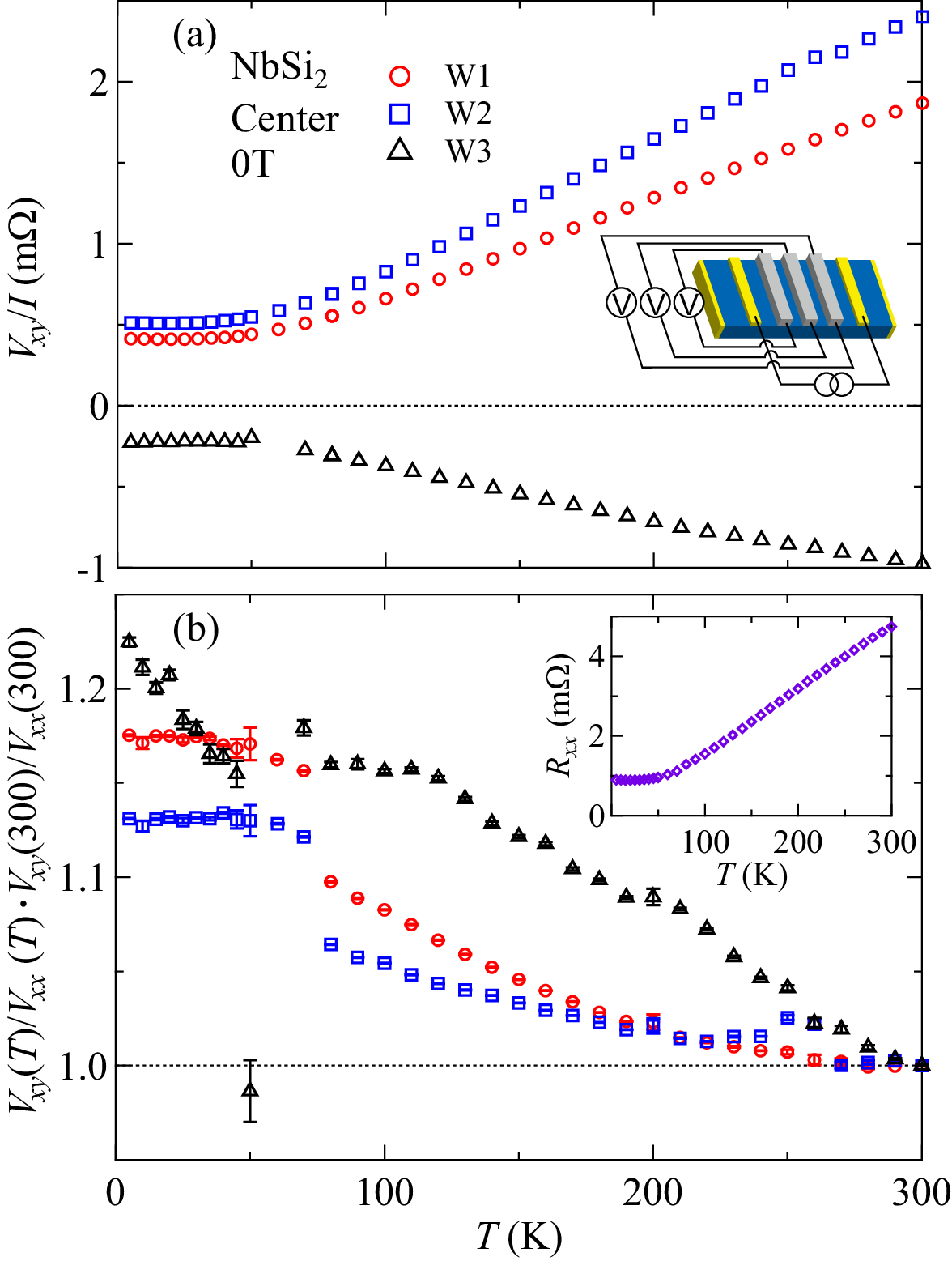}
\caption{Temperature dependence of (a) CISS signals $V_{xy}/I$ in the ``center'' region, and (b) signal ratio of $V_{xy}$ and $V_{xx}$ normalized at 300\,K for bulk polycrystalline NbSi$_2$ with the three W detection electrodes. The inset in (b) shows the temperature dependence of the longitudinal resistance $R_{xx}$.     
}
\label{temp_dep}
\end{figure}

A temperature dependence of the longitudinal resistance $R_{xx}$ and CISS signals $V_{\rm xy}$ was examined with the polycrystal NbSi$_2$ with the three W electrodes, shown in Fig.~6(a). 
$R_{xx}$ was measured with the standard four-probe method, in which the current was applied through the electrodes attached to the side of sample and the longitudinal voltage $V_{xx}$ was detected between the inner electrodes. 
$V_{xy}$ was measured at the three W electrodes in the ``center'' region, where $R_{xx}$ was measured, as schematically shown in the inset of Fig.~S\ref{temp_dep}(a).

$V_{xy}$ observed at the W3 detection electrode exhibits the opposite polarity to those at the W1 and W2 electrodes even though the same electrodes are used for the charge current application.  

These results exclude the contribution of nonuniform charge currents due to local resistance changes in the crystal and electrodes. Rather, the results suggest that the polarity of transverse signals reflects a distribution of the handedness under the detection electrodes and thus are due to the CISS effect. 

$V_{xy}$ decreases with reducing  temperature and reaches the residual values, as shown in Fig.~S\ref{temp_dep}(a).
$R_{xx}$, shown in the inset of Fig.~S\ref{temp_dep}(b), exhibits a typical temperature dependence of normal metals.
The signal ratio of $V_{xy}/V_{xx}$, which is normalized at 300\,K, is shown as a function of temperature in Fig.~S\ref{temp_dep}(b). The ratio increases with decreasing temperature, clarifying different temperature dependences of $V_{xy}$ and $V_{xx}$. 
This observation indicates that the present transverse voltage is not due to misalignments of the detection electrodes but to the intrinsic CISS effect.

\end{document}